\documentclass[prl,amssymb,twocolumn,floatfix,letter]{revtex4}
\usepackage{graphicx}
\begin{document}

\title{Multiscaling, Ergodicity and Localization in Quasiperiodic Chains}

\author{ V.~Z.~Cerovski }
\author{ M.~Schreiber }
\affiliation{ Institut f\"ur Physik, Technische Universit\"at, D-09107
Chemnitz, Germany. } 
\author{U.~Grimm}
\affiliation{ The Open University, Applied Mathematics Dept., Milton Keynes,
MK7 6AA, UK. } \date{\today}
\begin{abstract}
We report results of numerical simulations of wave-packet dynamics in a
class of chains consisting of two types of weakly coupled clusters
arranged in a quasiperiodic sequence. Properties of eigenstates are
investigated using perturbation theory of degenerate levels in the
coupling strength $v$ and by numerical diagonalization.  Results show
that wave packets anomalously diffuse {\it via}\/ a two-step process
of rapid and slow expansions, which persist for any $v>0$.  An
elementary analysis of the degenerate perturbation expansion reveals
that non-localized states may appear only in a sufficiently high order of
perturbation theory, which is simply related to the combinatorial
properties of the sequences. Numerical diagonalization furthermore
shows that eigenstates ergodically spread across the entire chain for
$v>0$, while in the limit as $v\to 0$ ergodicity is broken and
eigenstates spread only across clusters of the same type, in
contradistinction with trivial localization at $v=0$.  An
investigation of the effects of a single-site perturbation on
wave-packet dynamics shows that, by changing the position or strength
of such an impurity, it is possible to control the long-time wave-packet
dynamics.  By adding a single impurity it is possible to induce
wave-packet localization on individual subchains as well as on the
whole chain.
\end{abstract}

\maketitle

Understanding the relation between spectral properties of a given
Hamiltonian and the dynamics of wave-packets governed by it remains one
of the elementary questions of quantum mechanics that still poses
significant challenges, further emphasized by the discovery of
quasicrystals~\cite{Shechtman84,Torres03}.  While spectra of many
Hamiltonians decompose into a point part and an absolute continuous
part with, respectively, bound (localized) and unbound (delocalized or
extended) eigenstates, there is a large variety of Hamiltonians whose
spectra, for certain values of the parameters, are neither pure point
nor absolute continuous nor a combination of both. In this case the
spectrum contains a singular continuous part, with multifractal
eigenstates and/or Cantor-set spectra. Examples are Harper's model of
an electron in magnetic field~\cite{HarpersModel}, the kicked
rotator~\cite{kickedrotator}, as well as the Anderson model of an
electron in disordered medium~\cite{Schreiber91}. In the case of an
electron in a quasiperiodic system, as studied in this Letter, many
examples lead to spectra which are purely singular
continuous~\cite{singularcontinuous}.

It is known that a particle's return probability decays with the same
powerlaw as the scaling of the local density of states~\cite{delta}, and
that the spreading of the wave-packet width $d(t)$ (defined below) exhibits
anomalous diffusion, $d(t)\sim t^{\beta}$, with $0<\beta < 1$ ($\beta=0$
corresponds to the absence of diffusion, $\beta=1/2$ to classical diffusion
and $\beta=1$ to ballistic spreading)~\cite{transport}.  In addition, the
wave-packet dynamics exhibits multiscaling, where different moments of the
wave-packet scale with different, nontrivially related exponents $\beta$
\cite{Guarneri,Evangelou93,Wilkinson94,Piechon96,Barbaroux01}.  While
wave-packet localization implies a pure-point spectrum, the converse is not
true, and the more refined notion of semi-uniform localization is
necessary~\cite{delRio95}.  The exact relations between particle dynamics
and singular or absolute continuous spectra, on the other hand, is less
well understood.  As a rule, systems with singular continuous spectra
exhibit anomalous diffusion, while absolute continuous spectra may lead to
either anomalously diffusive or ballistic dynamics~\cite{Zhong95}.

In this Letter, we study eigenstate and wave-packet properties of an
electron moving in a one-dimensional quasiperiodic system described by the
tight-binding Hamiltonian $H=\sum_{\langle j
j'\rangle}(t_{jj'}c_j^\dagger c_{j'}+\mbox{H.c.})$. Here $c_j
(c_j^\dagger)$ is the annihilation (creation) operator of an electron
at site $j$, and the nearest-neighbors-only hopping integrals
$t_{jj'}$ take values $1$ or $0\le v\le 1$ for, respectively, letters
$l$ or $s$ (``large'' and ``small'' hoppings) of a quasiperiodic
sequence of letters obtained by the inflation rules $\{s\mapsto l,
l\mapsto lsl^{n-1}\}$ iterated $m$ times, starting with the letter
$s$.  Denoting the length of the $m$-th iterant by $N_m$, these
sequences have the property that $N_m/N_{m+1}\to\delta_n$ as
$m\to\infty$, where $\delta_n$ is an irrational number with continuous
fraction representation $[n,n,n,\dots]$.  The quantities of interest
are the solutions of the eigenproblem $H\psi_k(j)=E_k\psi_k(j)$, $k=1,\dots
N_m+1$, as well as the width $d(t)$ of a wave-packet $\Psi(j,t)$,
\begin{equation} 
d^2(t) = \sum_{j=1}^{N_m+1}|j-j_0|^2\;\;|\Psi(j,t)|^2,  
\end{equation}
which is initially localized at a site $j_0$, so
$\Psi(j,t=0)=\delta_{j,j_0}$,

Figure~1 shows the change of the wave-packet width with time for the
silver-mean ($n=2$) model.  While the asymptotic behavior can be
characterized as anomalous diffusion $d(t)\sim t^{\beta}, 
\beta\approx 0.2$, there are also ``flat'' parts in the regime of 
strong quasiperiodic modulation (small $v$).  As demonstrated by the 
insets in Fig.~1, these can be characterized by the existence of time
intervals, exponentially growing at a constant rate for a given
$v$, during which $d(t)$ strongly oscillates in a self-similar manner
reflecting the hierarchical structure of system, but nevertheless 
remains bounded from above by a constant.  Further inspection 
of the wave-packet behavior reveals that breathing modes are
responsible for the oscillations, while the wave-packet spreading
itself is limited to low-amplitude ``leaking'' out of the region
in which it is confined. Eventually, the wave-packet expands fast to 
reach the next level (this spreading can be described as 
$d(t)\propto t^{\beta'}$ with $\beta'\approx 0.79$), and the whole 
process repeats.

Such a behavior is in an agreement with results of
Ref.~\cite{Wilkinson94}, where, based on a qualitative model of the
wave-packet spreading in the semiclassical approximation and numerical
simulations, it was argued that an hierarchical splitting of the
spectrum into constant-width bands leads to a step-like behavior with
$\beta'=1$, which then is smoothed out due to the (broad) distribution
of band widths.  The here observed value $\beta'<1$ for small $v$ suggests
that there is a dispersion of band widths even in the limit as $v\to
0$ (when band widths go to 0~\cite{Passaro92}), and that the
self-similar spreading of the wave-packet is only an approximate
description of a more general multiscale dynamics.  This
self-similarity can be used, {\it e.g.}, for calculating $\beta$ by
$\beta\approx\delta\ln d/\delta\ln t$, where $\delta\ln d$ and
$\delta\ln t$ are the horizontal and vertical displacements 
between each of the two steps in Fig.~1, respectively, giving in
the case studied $\beta\approx 0.20$ for $v=0.025$.

\begin{figure}
\includegraphics*[width=3.0in]{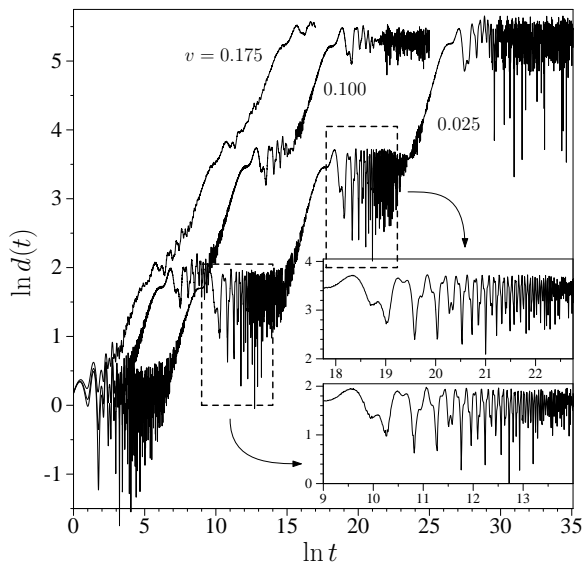}
\caption{Evolution of the width $d(t)$ of a wave-packet initially 
localized in the middle of the sliver-mean chain for $m=9$, for several 
small values of $v$.  The insets show two magnified ``steps'' for 
$v=0.025$.}
\end{figure}

The self-similarity of quasiperiodic sequences was previously used in
a renormalization-group perturbative expansion that provided a great
deal of insight into the eigenstate properties~\cite{Niu86}, and
showed multiscaling of wave-packet dynamics~\cite{Piechon96}.  Here
we focus on ergodic rather than hierarchical~\cite{Niu90} properties.
By an elementary analysis of the perturbation theory of degenerate
levels at $v=0$ for small $v$ we show that (a) for any $v>0$
eigenstates delocalize, in contradistinction to (trivial) localization
at $v=0$; and (b) in the limit as $v\to 0$ eigenstates delocalize across
only one set of clusters containing the same number of atoms (subcluster
localization due to ergodicity breaking).

Raleigh-Schr\"odinger theory allows the recursive construction of
matrices in subspaces of a given degenerate eigenenergy to a given
order $p$, whose diagonalizations (the ``secular problem'') yield
corrections to the unperturbed eigenenergies, hopefully accurate up to
$O(v^{p+1})$. In the case of an unperturbed chain ($t_{jj'}=t$), for
instance, the solution of the secular problem in first order in $t$ is
equal to the exact solution of the problem.

For the quasiperiodic sequences considered here, however, such
perturbation expansions yield two qualitatively different types of
solutions, depending on the values of $p$ and $n$.  To that end we
first notice that approximants for a given $n$ consist of two types of
words, $l^{n-1}$ and $l^n$, separated by the letter $s$, corresponding
to clusters with $n$ and $n+1$ atoms, respectively, coupled
{\it via} hopping of strength $v$ which is treated as a
perturbation.  The unperturbed system then has $2n+3$ degenerate
levels with all eigenstates localized on individual clusters.  In
higher orders of perturbation theory, these localized states spread as
the coupling among the clusters of the same type is taken into
consideration, and, for a sufficiently high order, delocalize across
the whole chain.  Since the maximal number of letters $s$ between two
consecutive clusters of length $k$ is also $k$, the eigenstates of the
clusters $l^{n-1}$ and $l^n$ delocalize only in the $n$ and $n+1$
order of the perturbation theory, respectively.  More precisely, 
the dimension of the secular problem for each cluster of length 
$q=n,n+1$ changes from at most $O(n^2)$ for $p<q$ to $O(N_m)$ 
for $p\ge q$.  Only the latter case allows for multifractal and/or 
extended states.

As an example we analyze the case $n=2$.  There are $7$ unperturbed
levels, with $E_{lll}^{(0)}=\pm({\sqrt 5}\pm 1)/2$ and
$E_{ll}^{(0)}=\pm{\sqrt 2},0$.  In first order, the only correction
are $6$ bands linear in $v$, splitting off the three $E_{ll}^{(0)}$
bands, because $sllslls$ is the only possibility where two clusters of
the same type are connected by a single $s$ bond.  For $p=2$, all $ll$
clusters become connected while $lll$ states are still not extended
due to the existence of $lllsllsllslll$ sequences in the chain. For
$p=3$, states belonging to $lll$ bands delocalize as well.

\begin{figure*}
\includegraphics*[width=6.4in]{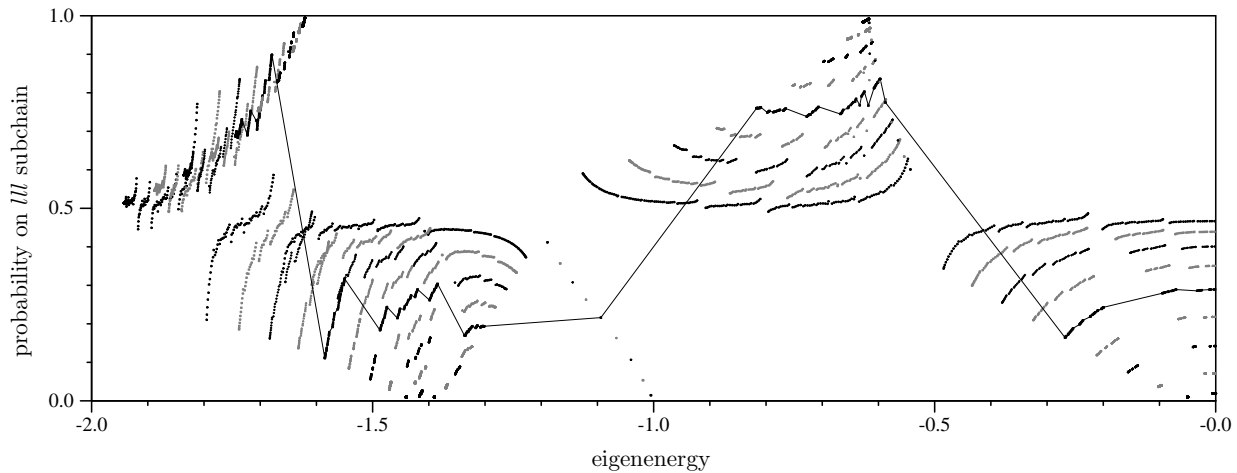}
\caption{Probability that a particle in an eigenstate with the given energy is
on the $lll$ sublattice, for $v=0.1,0.2,\dots 0.9$ (going from top to
bottom in the central part of the plot). The line connecting points for $v=0.5$
is for guidance only. Only states $E<0$ are shown due to the symmetry about
the band center.}
\end{figure*}

To investigate further the issue of convergence, we notice that, even when
calculated to all orders of $v$, the secular problems for the two types of
clusters give solutions that are inevitably localized on the clusters of the
given type and with zero component on the cluster of the other type.  For
various values of $v$, we check  by numerical diagonalization whether this is
confirmed. Fig.~2 shows $p_{lll}(E_k)\equiv\sum_{j\in lll} |\psi_k(j)|^2$, the
total probability that the particle in an eigenstate $\psi_k$ with an energy
$E_k$ will be on the cluster $lll$.  The result implies that $p_{lll}$ strongly
depends on the energy band of the given eigenstate for small $v$, being large
for the states belonging to the $lll$ bands, and vanishing for the states from
$ll$ bands in the limit $v\to 0$.  While Fig.~2 shows results for $m=9$, we
have also checked $p_{lll}$ for the iterants $m=10,11,12$ to see whether there
is any systematic deviation from the values shown in Fig.~2. We found that,
while there are many more states for larger $m$'s, the values of $p_{lll}$ do
not shift systematically for almost all of the states . Instead, the additional
points cluster in the same way as in Fig.~2.

\begin{figure*}
\includegraphics*[width=6.0in]{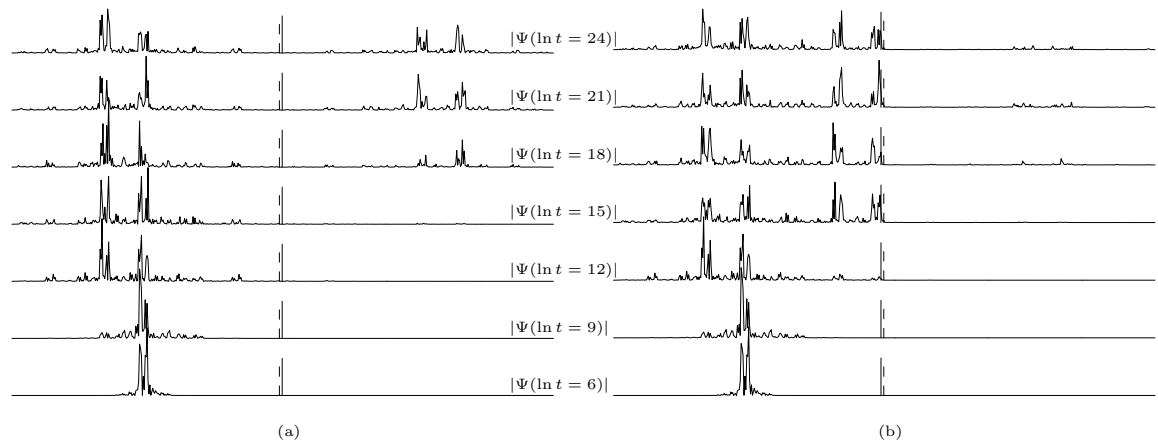}
\caption{Snapshots of the evolution of two wave-packets, both starting
as $\delta$-functions in a local environment $\dots sl(x)lls\dots$ at
the same initial site $(x)$, in the presence of an impurity. In the 
two panels (a) and (b), the impurity is located in a local environment
$\dots sll(b)sll(a)slls\dots$ at the position marked by (a) and (b), 
respectively, as indicated in each panel by a vertical line. 
For easier comparison, the vertical dashed line in each panel marks the
position of the impurity in the other panel.  The long-time wave-packet
dynamics exhibit high sensitivity on whether the impurity is located or not
on the cluster of the same type as the one where the wave-packet was localized
initially. The remaining parameters are $n=2$, $m=8$, $v=u=0.1$.}
\end{figure*}

This supports the possibility that eigenstates remain ergodic ({\it
i.e.}, spread over both types of clusters) even for very small values
of $v$, and that, in turn, the Schr\"odinger-Raleigh perturbation
expansion of degenerate levels has zero radius of convergence but,
nevertheless, might still be accurate in the limit as $v\to 0$ since
$p_{lll}(v\to 0)\to 0$ or 1 (which is, however, only a necessary but not a
sufficient condition).  On the other hand, the nearly non-ergodic
behavior of wave-packets for small $v$ has interesting consequences when
a single impurity is added, as we show next.

An impurity $H'=u_{i} c_{i}^\dagger c_{i}$ added at a site $i$ will, for large
$u$ ({\it e.g.}, $u\gg 1$), act as a barrier, effectively cutting the chain
into two halves.  The consequence of ergodicity in this limit is that the
wave-packet reflects off the impurity independently of its initial site, even
if $v$ is small.  For $u\to 0$, on the other hand, the unperturbed wave-packet
propagation of the case $u=0$ is restored.  Understanding the wave-packet
propagation in the regime of intermediate $u$ values, however, still poses
significant challenges and surprising results~\cite{delRio95}.  We performed
several numerical experiments for various initial positions of impurities and
wave-packets for intermediate values of $u$, and one common situation is shown
in Fig.~3.  The evolution of an initially localized wave-packet exhibits high
sensitivity on the position of the impurity, approaching two quite different
stationary states.  In particular, while in Fig.~3(a) the final state is just
slightly perturbed from a final state for $u=0$, in Fig.~3(b) certain
parts of the wave packet get reflected, with a small-amplitude ``leak''
reminiscent of the similar process mentioned above.  This kind of wave-packet
dynamics is a consequence of the nearly non-ergodic properties of eigenstates
discussed above (cp.~Figure~2).

\begin{figure}
\includegraphics*[width=3.1in]{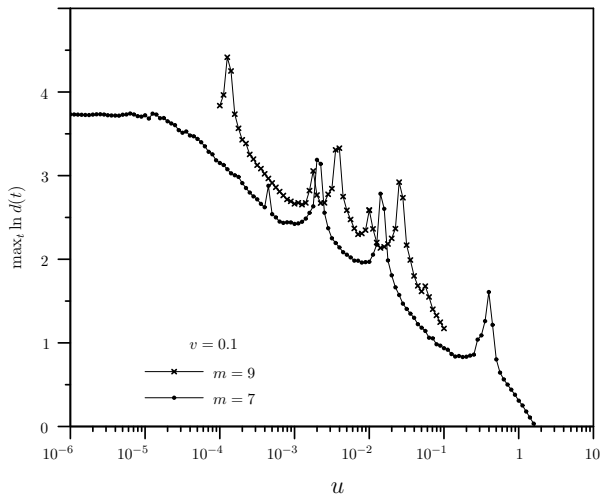}
\caption{Maximal width of a wave-packet attained during its evolution in the
presence of a single impurity of strength $u$ located at the initial site of
the wave-packet.}
\end{figure}

Finally, we address the influence of an impurity at the initial site of the
wave-packet on its dynamics, by studying the dependence of the final
wave-packet width on the value of $u$.  Fig.~4 shows the maximum value of
$d(t)$ attained in the course of the evolution of a wave-packet, which was
initially localized at the site $j_0=N_m/2$, in the presence of an impurity
located at the same site.  The $m=7$ iterant of the silver-mean model has
unperturbed wave-packets exhibiting one full step from Fig.~1, reaching the
final stationary state after beginning of the second step.

Fig.~4 shows that, apart from the strongly localized final wave-packets
for large $u$ and unperturbed final stationary wave-packets for small
$u$, there is a wide range of values of $u$ for which the final width
of the wave-packet is significantly reduced even for $u\ll v$,
signaling (dynamical) localization.  There are nevertheless several
well defined peaks about some values of $u$ where the maximum
wave-packet width is significant enhanced compared with the case for
slightly smaller and slightly larger values of $u$.  We additionally
checked whether such peaks persist when the system size is increased, by
performing the same calculation for the $m=9$ iterant. Fig.~4 shows
that there is again a peak structure, with both small and large peaks.
The relation of the values of $u$ where these peaks appear and the
spectrum of $H$ remains unclear.  Repeating similar numerical
experiments for various values of $v$, we find that the peak structure
persists as long as the band widths are smaller than the band gaps
($v\lessapprox 0.4$ in the silver-mean case). These results, however,
show that in quasiperiodically modulated quantum wires one can
strongly influence the electronic transport by inducing local
perturbations that act as sort of control gates for long-range quantum
interference effects in electronic transport.

\end{document}